\documentclass[showpacs,twocolumn,amsmath,amssymb,prb]{revtex4-1}

\usepackage{graphicx}
\usepackage{amsmath,amsfonts,amssymb,amsthm}
\usepackage{fancyhdr}
\usepackage{mathrsfs}

\begin{document}

\title{Fragility of multi-junction flux qubits against quasiparticle tunneling}

\author{Juha Lepp\"akangas and Michael Marthaler}

\affiliation{Institut f\"ur Theoretische Festk\"orperphysik
and DFG-Center for Functional Nanostructures (CFN), Karlsruhe Institute of Technology, D-76128 Karlsruhe, Germany}


\begin{abstract}
 We study decoherence in superconducting qubits due to quasiparticle tunneling which
 is enhanced by two known deviations from the equilibrium BCS theory.
 The first process corresponds to tunneling of an already existing quasiparticle across the junction.
 The quasiparticle density is increased, e.g., because of an effective quasiparticle doping of the system.
 The second process is quasiparticle tunneling by breaking of a Cooper pair.
 This can happen at typical energies of superconducting qubits if there is an extended quasiparticle density
 inside the gap. We calculate the induced energy decay and pure dephasing rates
 in typical qubit designs. Assuming the lowest reported value of the non-equilibrium quasiparticle density in Aluminum,
 we find for the persistent-current flux qubit decay times of the order of recent measurements.
 Using the typical sub-gap density of states in Niobium we also reproduce observed decay
 times in the corresponding Niobium flux qubits.
\end{abstract}


\maketitle



\section{Introduction}

 A basic building block of a superconducting quantum bit is the Josephson junction, which allows
 coherent Cooper-pair tunneling. This Josephson current provides
 the necessary non-linear element in an electric circuit and
 enables reduction of the externally controllable quantum dynamics to involve only two eigenstates~\cite{ShnrimanSchoenReview}.
 A drawback for superconducting qubits has been that these systems can be very sensitive
 to various effects in their nearby environments.
 The influence of two major decoherence sources, charge fluctuations and two-level fluctuators,
 have been substantially reduced in recent years. Qubits that are less sensitive to changes in background charge have been build
 \cite{JensKochTransmon,Fluxonium}, while fluctuators have been removed by decreasing the junction size.
 Recently, transmons and persistent-current flux (p-flux) qubits with extremely long decay times
 have been demonstrated \cite{SuperTransmon,NakamuraSuperFux} .
 It is therefore of great interest to investigate mechanisms of decoherence that were previously unobservable.
 Temperature dependent decay time measurements in phase\cite{MartinisQP1} and
 transmon qubits\cite{QPs_in_Transmons} suggest that non-equilibrium quasiparticles
 might have now become the main factor limiting the decay times of superconducting qubits.

 In this work we analyze qubit decay and dephasing due to quasiparticle tunneling. As the source of tunneling
 we consider the two following deviations from the equilibrium BCS state of superconductors.
 The first one is a non-equilibrium distribution of quasiparticles, which corresponds to a
 finite quasiparticle density above the gap, present at all temperatures~\cite{Shaw,Palmer,Pekola,GlazmanDevoretQPTheory,GlazmanDevoretQPTheoryLong}.
 Non-equilibrium quasiparticles
 could originate in a quasiparticle diffusion from higher temperature regions in an experiment,
 or through stray radiation, and are relatively long lived due to slow quasiparticle recombination rate~
 \cite{OwenScalapino,MartinisPRBDecayinQubits,MartinisStrayLight,CatelaniQPDissipation}.
 The second process we consider is quasiparticle tunneling by breaking of a Cooper pair, which can occur
 at typical energies of superconducting qubits if there is an extended quasiparticle density inside the gap.
 It has long been observed that the number of states inside the gap is vastly larger than the predictions of BCS theory~\cite{Likharev}.
 This effect is especially pronounced in Niobium, a material that is widely used in the production  of qubits~\cite{NiobiumQubits_DWAVE}.
 Nonetheless, Aluminum based superconducting devices are observed to posses sub-gap states too~\cite{Pekola,PekolaEE}, but with considerably smaller density
 as Niobium.

\begin{table}[b]
\begin{tabular}{|l|c|c|}
 \hline
 Qubit type & Non. QP (Al) &  Sup-gap DOS (Nb)\\ \hline
 Transmon\cite{transmon} & 1.2~ms  &  20~$\mu$s  \\
 Phase\cite{MartinisQP1} & 0.6~ms & 14~$\mu$s \\
  Capacitively-shu.~flux\cite{Capacativelyshuntedfluxqubti} & 80~$\mu$s & 3~$\mu$s\\
  Persistent-current flux\cite{pflux} & 12~$\mu$s & 0.3~$\mu$s \\ \hline
\end{tabular}
\caption{Comparison of decay times obtained by assuming a non-equilibrium
quasiparticle density $n_{\rm qp}=0.033/\mu$m$^3$ (second column, Aluminum) or
a Dynes parameter $\gamma_{\rm Nb}=\Gamma_{\rm D}/\Delta=10^{-2}$ (third column, Niobium).
We used qubit parameters specified in each of the citations at each qubit. We also
assume $\Delta_{\rm Al}=\Delta_{\rm Nb}/7=180$~$\mu$eV and the validity of the Ambegaokar-Baratoff
relation. Typical sub-gap density of states
in Aluminum ($\gamma_{\rm Al}\sim 10^{-4}-10^{-7}$) does not limit the coherence here.
}\label{tab_comparing_decay_times}
\end{table}

 Our results are that the two quasiparticle processes result in a similar type of qubit decay rate. This
 provides a clear comparison between the magnitudes of decoherence due to the two sources.
 For conventional Aluminum qubits non-equilibrium quasiparticles should have a more significant
 impact on decoherence,
 for Niobium qubits the sub-gap density of states.
 We also find that the sensitivity to quasiparticle tunneling
 is vastly different for differing qubit designs.
 By using the lowest reported density for non-equilibrium quasiparticles 
 in Aluminum~\cite{Pekola} we obtain p-flux qubit decay times similar to recent experiments \cite{NakamuraSuperFux},
 whereas for other qubit types
 the results are from two to three orders of magnitude greater
 than the observed ones.
 For Niobium  p-flux qubits we also reproduce usual experimental decay times by assuming a Dynes-type of
 quasiparticle sub-gap states, observed in various other experiments \cite{Toppari}.
 These central results are summarized in Table~\ref{tab_comparing_decay_times}.

 The paper is organized as follows. In section \ref{Sec_Foundations} we introduce the Hamiltonian and methods describing the qubit-quasiparticle dynamics.
 The section \ref{Sec_DecoherenceRates} is devoted to discussion of decoherence processes and the corresponding rate equations.
 We also discuss decay rates for single-junction qubits obtained by using typical experimental parameters of the quasiparticle environments.
 In section \ref{sec_Multi_Junction_Flux_Qubit} we generalize the treatment to multi-junction qubits,
 where a special concentration is given to flux qubits in their different designs.
 In section \ref{Sec_Conclusion} we summarize the results.


\section{Foundations}\label{Sec_Foundations}

Our starting point is the division of the total Hamiltonian into three parts,
\begin{eqnarray}\label{TotalHamiltonian}
H_{\rm total}=H_{\rm qubit}+H_{\rm qp}+H_{\rm T}.\label{eq_HamiltonianTotal}
\end{eqnarray}
Here the qubit Hamiltonian $H_{\rm qubit}$ describes the collective degrees of freedom that constitute our two-level system.
The quasiparticle Hamiltonian $H_{\rm qp}$ describes the electronic degrees of freedom in the superconductors,
treated independently of the collective part. Interaction between these two parts emerges
due to single-electron tunneling modeled by the tunneling Hamiltonian $H_{\rm T}$.
It describes, e.g., quasiparticle tunneling across
the Josephson junction which can cause transitions between the eigenstates of the qubit.
In the following we discuss in more detail the central properties of each of the three parts.

\subsection{Single-junction qubit Hamiltonian}
As the general single-junction qubit Hamiltonian we use the following sum of capacitive, inductive, and Josephson coupling energy
\begin{eqnarray}\label{QubitHamiltonian}
H_{\rm qubit}=E_CN^2+\frac{E_L}{2}(\varphi-\varphi_{\rm ext})^2-E_{\rm J}\cos\varphi.\label{eq_HamiltonianQubitSingle}
\end{eqnarray}
Here $N$ is the number operator of electron charges on the junction capacitor $C$. It is a conjugated variable
to the phase difference across the junction $\varphi$, satisfying $[\varphi,N/2]=i$.
This choice of qubit Hamiltonian corresponds to the phase and dc-flux qubits, i.e.,
 when the junction is placed in a superconducting loop with inductive energy $E_L$ and thread by flux $\varphi_{\rm ext}$.
For the transmon one has $E_{L}=0$ and periodic commutation relations $[N/2,e^{\pm i\varphi}]=\pm e^{\pm i\varphi}$,
related to the $2e$-quantization of the charge that tunnels across the Josephson junction. All qubits we consider work in the limit
where the Josephson coupling energy $E_{\rm J}$ is larger than the single-electron charging energy $E_C=e^2/2C$.
Generalization to multi-junction qubits is straightforward, see section~\ref{sec_Multi_Junction_Flux_Qubit}.

\subsection{Quasiparticle degrees of freedom}
 The quasiparticle degrees of freedom in superconducting leads are modeled by the Hamiltonian $H_{\rm qp}$.
 For all further calculations
 the quasiparticle properties are assumed to be independent of the qubit state.
 We consider here two models as two separated causes of quasiparticle tunneling at energies
 well below the superconducting gap $\Delta$. We label these as model $I$ and model $II$.

 Signatures of the electronic structure in the two models are found
 from correlation functions of type,
 \begin{equation}
 \begin{split}\label{eq_GreenDefined}
 G^{>}_{\alpha\alpha}({\bf k},t)&=\left\langle c_{{\bf k}\alpha}(t)c_{{\bf k}\alpha}^{\dagger}\right\rangle ,\,\, G^{<}_{\alpha\alpha}({\bf k},t)=\left\langle c_{{\bf k}\alpha}^{\dagger}c_{{\bf k}\alpha}(t)\right\rangle , \\
 F_{\alpha\beta}^{>}({\bf k},t)&=\left\langle c_{{\bf k}\alpha}^{\dagger}(t)c_{-{\bf k}\beta}^{\dagger}\right\rangle, F_{\alpha\beta}^{<}({\bf k},t)=\left\langle c_{{\bf k}\alpha}c_{-{\bf k}\beta}(t)\right\rangle.
 \end{split}
 \end{equation}
Here $c_{{\bf k}\alpha}^{(\dagger)}$ is an electron annihilation (creation) operator of the state $\bf k$ with spin $\alpha$ ($\alpha\neq\beta$).
The Fourier transforms of the correlation functions are related to spectral and distribution functions as~\cite{Barone}
 \begin{equation}\label{eq_FourierTransforms}
 \begin{split}
G^{\gtrless}_{\alpha\alpha}({\bf k},\omega)&=A({\bf k},\omega)f^{\pm}(\omega),\\
F^{\gtrless}_{\uparrow\downarrow}({\bf k},\omega)=-F_{\downarrow\uparrow}^{\gtrless}({\bf k},\omega)&=B({\bf k}.\omega)f^{\pm}(\omega).
\end{split}
\end{equation}
Here $f^-=f$ and $f^+=1-f$. In equilibrium $f=f_{\rm eq}(\omega)=1/[1+\exp(\omega/k_BT)]$.
We define the normalized density of states $n$ and the pair density $p$ as
 \begin{equation}\label{eq_Densities}
 \begin{split}
 n(\omega)&=\frac{1}{\pi D}\sum_{{\bf k}}A({\bf k},\omega),\\
 p(\omega)&=\frac{1}{\pi D}\sum_{{\bf k}}B({\bf k},\omega).
 \end{split}
 \end{equation}
 Here $D$ is the density of states nearby the Fermi surface, including spin.
These functions, added with the distribution function $f$, have their own characteristic forms in the two models and are all we need to know
about the quasiparticle environments in the final forms of the qubit decoherence rates (see section \ref{Sec_DecoherenceRates}).

\subsubsection{Model $I$: Non-equilibrium quasiparticles}
 In model $I$ we assume that in each lead quasiparticles can be described by the BCS Hamiltonian
 but exist with a general (non-equilibrium) probability $f$.
 In the BCS model the quasiparticle and pair densities have the form
 \begin{equation}\label{broadening}
 \begin{split}
 n(\omega)&=\frac{\omega }{\sqrt{\omega^2-\Delta^2}}\Theta(\omega^2-\Delta^2){\rm sgn}(\omega), \\
 p(\omega)&=\frac{\Delta}{\sqrt{\omega^2-\Delta^2}}\Theta(\omega^2-\Delta^2){\rm sgn}(\omega) .
  \end{split}
 \end{equation}
An important quantity is the total density of quasiparticles
 \begin{equation}\label{eq_n_qp_defined}
 n_{\rm qp}=2D\int_{\Delta}^{\infty}n(\omega)f(\omega)d\omega.
 \end{equation}
 In calculations $n_{\rm qp}$ is assumed to be a given constant.
 This model has also been investigated in Refs.~\onlinecite{GlazmanDevoretQPTheory,GlazmanDevoretQPTheoryLong}.

\subsubsection{Model $II$: Sub-gap density of states}
 As the other source of quasiparticles we consider
 the presence of quasiparticle states below the gap.
 We assume that the densities can be expressed in the same form as for the BCS theory, but with the following
 broadening,
 \begin{equation}\label{broadening2}
 \begin{split}
 n(\omega)&={\rm Re}\left\{ \frac{\omega+i\Gamma_{\rm D}}{\sqrt{(\omega+i\Gamma_{\rm D})^2-(\Delta+i\Delta_2)^2}}\right\}{\rm sgn}(\omega), \\
 p(\omega)&={\rm Re}\left\{ \frac{\Delta+i\Delta_2}{\sqrt{(\omega+i\Gamma_{\rm D})^2-(\Delta+i\Delta_2)^2}}\right\}{\rm sgn}(\omega).
 \end{split}
 \end{equation}
 Here $\Gamma_{\rm D}\ll\Delta$ is the so-called Dynes parameter~\cite{Dynes} and $\Delta_2\ll\Delta$ is a possible imaginary part
 of the superconducting gap~\cite{Barone}. We consider $\Gamma_{\rm D}$ and $\Delta_2$
 to be given parameters and independent of energy.
 The effect of these modifications of the BCS theory are very similar nearby the gap,
 but differ at low energies $\omega\approx 0$ (see section \ref{sec_LowEnergyApproximation}).
 The broadening through an imaginary part of the
 superconducting gap~\cite{ConfSeries}, $\Delta_2$ ($\Gamma_{\rm D}=0$), is supported by the microscopic Eliashberg theory.
 However, a constant imaginary part of the energy, $\Gamma_{\rm D}$ ($\Delta_2=0$), 
 usually fits better to experiments done with, e.g., Niobium~\cite{Toppari,Dynes,Pashkin}.
 For all further purposes we will only consider finite $\Gamma_{\rm D}$ and $\Delta_2=0$.

\subsection{Quasiparticle tunneling}
The interaction between the qubit and the quasiparticle environments occurs due to the possibility of
quasiparticles to tunnel across the junction(s)
and simultaneously change the charge configuration in the qubit space.
This is described by the {\em modified} tunneling Hamiltonian $H_{\rm T}$
\begin{eqnarray}\label{Ham4}
H_{\rm T}=\bar H_{\rm T}+E_{\rm J}\cos\varphi, \label{eq_QuasiparticleTunneling1}
\end{eqnarray}
where the {\em bare} tunneling Hamiltonian has the form
\begin{eqnarray}\label{Ham4}
\bar H_{\rm T}=t\sum_{{\bf kl}\alpha}(c_{{\bf k}\alpha}^{\dagger}c_{{\bf l}\alpha}\hat T+{\rm h.c.}). \label{eq_QuasiparticleTunneling2}
\end{eqnarray}
Here the states $\bf k$ and $\bf l$ belong to the opposite sides of the
junction. We assume a constant tunneling matrix element $t$, related to the normal state tunnel resistance as $R_{{\rm T}}=\hbar/ t^2  D^2\pi e^2$.
The charge-transfer operator $\hat T=\sum_N\vert N+1\rangle\langle N\vert= e^{i\varphi/2}$
accounts for the corresponding changes in the charge number of the leads.
Cooper-pair tunneling, which is a second-order process in quasiparticle
tunneling, is already included in the qubit Hamiltonian in an approximative way as the operator
$-E_J\cos\varphi$. Therefore
we need to subtract it from $\bar{H}_{\rm T}$ to avoid double counting.

\section{Decoherence rates}\label{Sec_DecoherenceRates}
The second-order expansion of the qubit's density-matrix equation of motion in the tunneling Hamiltonian leads to contributions describing
qubit decoherence and parameter renormalization. Here we
discuss in detail the decoherence terms, i.e., terms leading to energy decay and pure dephasing. We then
estimate their magnitude for usual single-junction qubits. An exact derivation of the rates,
a formulation for the energy-level renormalization effects, 
and a discussion for the effect of higher-order terms is given in the Appendix.

The system we consider is a two-level system interacting with a fermionic bath. The
coupling to the bath can be divided
into two parts: (i) the part causing transitions, $\propto\sigma_x$, and (ii) the part causing pure dephasing, $\propto\sigma_z$.
The decay rate, $\Gamma_1$, is a result of coupling to $\sigma_x$. This leads also to the ordinary dephasing rate $\Gamma_2=\Gamma_1/2$.
Fluctuations through $\sigma_z$ lead to the pure dephasing rate $\Gamma_{2^*}$,
which can be interpret to be a result of low-frequency fluctuations in the qubit energy splitting due to coupling to the bath~\cite{ShnrimanSchoenReview}.

\subsection{Tunneling processes}\label{processes}
In leading order we obtain two distinct types of tunneling processes causing decoherence.
The first one is tunneling of
an existing quasiparticle. In the case of the BCS state this is described by an operator of the type
\begin{equation}\label{eq_processI}
\hat{O}_{1}=u_{\bf{k}}u_{\bf{l}} \hat T - v_{\bf{k}}v_{\bf{l}}  \hat T^{\dagger}.
\end{equation}
This process contributes mainly in model $I$.
The second one is the breaking of a Cooper pair, described by the BCS operator of the type
\begin{equation}\label{eq_processII}
\hat{O}_{2}=u_{\bf{k}}v_{\bf{l}} \hat T + v_{\bf{k}}u_{\bf{l}}  \hat T^{\dagger}.
\end{equation}
This process contributes only in model $II$.
Important here is that the operators $\hat O_i$ are superpositions of two electron-tunneling directions, $\hat T$ and $\hat T^{\dagger}$,
as long as $u,v\neq 0$. 
As the qubit states are superpositions of different charge states,
the direction of electron tunneling becomes indistinguishable and the two tunneling processes  interfere.
This gives rise to a phase dependence in
quasiparticle tunneling, similar to the $\cos\varphi$-term in the classical Josephson effect~\cite{JuhaQPTunnelingPRB}.
This interference effect has also an important role in decoherence of superconducting qubits, as discussed below.


\subsubsection{Qubit decay}
For the qubits considered in this work the qubit decay rate can be written in the form
 \begin{equation}\label{eq_Gamma12_defined}
 \begin{split}
\Gamma_{1}&=\frac{2}{R_{{\rm T}}e^2} \left\vert \langle \uparrow |\hat T| \downarrow \rangle \right\vert^2 \int_{-\infty}^{\infty}d\omega\int_{-\infty}^{\infty}d\omega'f^-(\omega)f^+(\omega') \\
\times&  \left [n(\omega)n(\omega')+p(\omega)p(\omega')\cos\varphi_0 \right]\delta(\omega-\omega'-\delta E).
\end{split}
\end{equation}
Here $\delta E$ is the energy-level splitting of the qubit and the angle $\varphi_0$ is defined as
\begin{eqnarray}\label{eq_AngleDefinition}
e^{i\varphi_0}=-\frac{\langle \uparrow |\hat T| \downarrow \rangle \langle \downarrow |\hat T| \uparrow \rangle}{\left\vert \langle \uparrow |\hat T| \downarrow \rangle \right\vert^2}.
\end{eqnarray}
The matrix elements depend on the choice of superconducting qubit, but as we will discuss, they are in fact very similar for broad classes of qubits.

Expressions such as (\ref{eq_Gamma12_defined}) are often derived in the BCS excitation picture, where integration
over only positive (excitation) energies emerges. In this semiconductor-type presentation negative energies
appear as well. But the result of the two presentations is the same.
The positive integration region of both frequencies $\omega$ and $\omega'$ corresponds to
tunneling of an existing quasiparticle to one direction, as the negative  integration region to the other region.
The contribution with positive $\omega$ but negative $\omega'$ describes breaking of a Cooper pair during
the single-electron tunneling. It can exist only if there is an extended states below the gap (because $\delta E<2\Delta$).
The contribution with positive $\omega'$ but negative $\omega$
contributes only if $\delta E<0$ and would correspond to recombination of two quasiparticles
with excitation of the qubit.

\subsubsection{Qubit decay in the low-energy approximation}\label{sec_LowEnergyApproximation}
To simplify equation (\ref{eq_Gamma12_defined}) we assume $\delta E\ll\Delta$
and that the width of the quasiparticle distribution ($\sim k_BT$) is much smaller
than the qubit splitting, $\delta E$.
This means that for model $I$ the distribution $f$ is nonzero above the gap ($\omega\geq\Delta$) practically only
in a very narrow region [we have always $f(-\omega)=1-f(\omega)=f^+(\omega)$]. On the other hand,
for the model $II$ with Dynes broadening (and $\Delta_2=0$) one can approximate
\begin{equation}
\begin{split}
n&= \frac{\Gamma_{\rm D}}{\Delta},\\
p&= \frac{\omega}{\Delta}\frac{\Gamma_{\rm D}}{\Delta}.
\end{split}
\end{equation}
For a finite imaginary part $\Delta_2$ and $\Gamma_{\rm D}=0$ one would have
$n= (\omega/\Delta)(\Delta_2/\Delta)$ and $p= (\omega/\Delta)^2(\Delta_2/\Delta)$.
However, in the following we assume a Dynes-type broadening \cite{Toppari,Dynes,Pashkin}
when considering model $II$.

In the discussed low-energy approximation we obtain the following common form for the decay rates,
\begin{equation}
\begin{split}\label{eq_MainResult}
\Gamma_1&=\frac{2\xi_iM_1^2}{R_{\rm T}e^2}\left(1+\epsilon_i\cos\varphi_0\right)\,\, ,\,i=I,II\, ,\\
M_1^2&=\left\vert \langle \uparrow |\hat T| \downarrow \rangle \right\vert^2.
\end{split}
\end{equation}
Important here is that the rates are proportional to $\xi_i$,
defined for the two cases as
\begin{equation}
\begin{split}
\xi_I &=\frac{n_{\rm qp}n(\Delta+\delta E)}{D}\, , \\
\xi_{II} &=\delta E\left(\frac{\Gamma_{\rm D}}{\Delta} \right)^2. \label{eq_MainDensity}
\end{split}
\end{equation}
The form $\Gamma_1\propto 1+\epsilon_i\cos\varphi_0$ for the decay rate (\ref{eq_MainResult}) is a result of the
interference effect in quasiparticle tunneling~\cite{JuhaQPTunnelingPRB} (see the discussion in section \ref{processes}).
Its magnitude and sign are given by
\begin{equation}
\begin{split}
\epsilon_I &=\frac{1}{1+\delta E/\Delta}\, , \\
\epsilon_{II} &=-\frac{1}{6}\left(\frac{\delta E}{\Delta}\right)^2.\label{eq_CosSign}
\end{split}
\end{equation}
For model $I$ we have $\epsilon_I\approx 1$ whereas
for model $II$ $\epsilon_{II}\approx 0$.
Physical interpretation for this is that the subgap states in model $II$
are close to metallic states and do not show significant pair correlations ($p\approx 0$), needed for the effect.

\subsubsection{Pure dephasing}
In the leading-order expansion of the qubit's time evolution the pure dephasing appears as an extra decay
of coherent oscillations through transition terms proportional to $\langle\uparrow \vert T^{(\dagger)}\vert\uparrow\rangle$
or $\langle\downarrow \vert T^{(\dagger)}\vert\downarrow\rangle$.
The pure dephasing rate is then similar to the rate~(\ref{eq_Gamma12_defined}) by setting  $\delta E=0$,
and changing the matrix element,
 \begin{equation}
 \begin{split} \label{eq_MainResult2}
  \Gamma_{2^*}&= \frac{M_2^2}{R_{\rm T}e^2} \int_{-\infty}^{\infty}d\omega\int_{-\infty}^{\infty}d\omega' f^-(\omega)f^+(\omega') \\
  &\times \left[ n(\omega)n(\omega')+p(\omega)p(\omega')\cos\varphi_{0^*} \right]\delta (\omega-\omega')\, , \\
 M_2^2&= \left\vert \langle\uparrow\vert \hat T \vert\uparrow\rangle-\langle\downarrow\vert \hat T \vert\downarrow\rangle\right\vert^2 .
 \end{split}
 \end{equation}
 Here we have assumed that
\begin{eqnarray}\label{eq_AngleDefinition2}
e^{i\varphi_{0^*}}=-\left(\frac{\langle \downarrow |\hat T| \downarrow \rangle}
{\left\vert \langle \downarrow |\hat T| \downarrow \rangle \right\vert}\right)^2=-\left(\frac{\langle \uparrow |\hat T| \uparrow \rangle}
{\left\vert \langle \uparrow |\hat T| \uparrow \rangle \right\vert}\right)^2,
\end{eqnarray}
 which is the case for most of the qubits considered in this paper.
 The only exception is the flux qubit away from its symmetry point, discussed more detailed in section \ref{sec_Multi_Junction_Flux_Qubit}.

 The value of the integration in Eq.~(\ref{eq_MainResult2}) depends on the distribution $f$, which depends on the model used.
 In model $II$ we assume the equilibrium distribution $f_{\rm eq}=1/[1+\exp(\omega/k_BT)]$. Then
 the pure dephasing corresponds to fluctuations originating from tunneling of thermalized (sub-gap) quasiparticles.
 For this case the integration can be easily evaluated giving $(\Gamma_{\rm D}/\Delta)^2k_BT$.
 This means that for small temperatures $k_BT\ll\delta E$ the pure dephasing rate
 is much smaller than the decay rate. 

 In model $I$  the situation is more difficult
 due to the logarithmic divergence of the energy integral at the energy gap.
 Similar singularities appear also in other properties of superconductors~\cite{Tinkham},
 but stay finite due to finite lifetime effects or gap anisotropy.
 It is now crucial, which type of broadening is assumed. If one introduces a small imaginary part $\Delta_2$
 and accordingly redefines $n_{\rm qp}$ in Eq.~(\ref{eq_n_qp_defined}),
 then the integral in Eq.~(\ref{eq_MainResult2}) remains small.
 As a result of this the pure dephasing rate  stays small compared to the energy decay rate.
 Other methods for circumventing the divergence can produce larger results for the integral.
 However, as we will discuss in the next section, pure dephasing is additionally suppressed by small tunneling matrix elements.


\subsection{Discussion for single-junction qubits}\label{DiscussionForSingleQubits}

 We will now discuss the qubit part of the decoherence rates
 in the case of single-junction qubits. Specifically we consider the phase qubit, the transmon, the
 dc-flux qubit, the strongly anharmonic phase qubit \cite{Zorin} and the fluxonium. Strictly speaking the
 fluxonium is not a single junction qubit, but we consider the integrated Josephson junction array simply as an
 inductance. For a detailed discussion of the Fluxonium including quasiparticle tunneling in the junction array
 see Ref.~\onlinecite{GlazmanDevoretQPTheoryLong}.
 The discussion can be structured along two major properties.
 One is the squared magnitude of the qubit matrix element in Eq.~(\ref{eq_MainResult})
 or of the difference between matrix elements in Eq.~(\ref{eq_MainResult2}),
 divided by the tunneling resistance $R_{\rm T}$.
 The second is the phase difference $\varphi_0$ in the interference ($\cos\varphi_0$) term.

 When it comes to estimating the matrix elements qubits fall broadly in two classes. One class of qubits have eigenstates
 located in a single local minimum of the qubit's potential energy,
  \begin{equation}
   U=\frac{E_L}{2}(\varphi-\varphi_{\rm ext})^2-E_{\rm J}\cos\varphi\,.
  \end{equation}
 The lowest eigenstates are then similar to that of a harmonic oscillator and they are symmetric or antisymmetric
 around the local minimum. Here the decay and dephasing elements are in a good approximation given by
 \begin{equation}
 \begin{split}\label{eq_ElementsAnharmonic}
 M_1^2 &= \frac{E_C}{\delta E},\\
 M_2^2 & = \left(\frac{E_C}{\delta E}\right)^2.
 \end{split}
 \end{equation}
 In this class we have clearly the transmon and the phase qubit. A qubit with a quartic potential
 as proposed in Ref.~\onlinecite{Zorin} (strongly anharmonic phase qubit) has the same matrix
 element and despite its rather unusual potential, the fluxonium
 belongs to this group as well for $\varphi_{\rm ext}=0$.
 The qubit that bucks the trend is the dc-flux qubit,
 which in fact is very similar to the fluxonium for $\varphi_{\rm ext}=\pi$.
 With its double well potential it is very different
 from all the other qubits. We will discuss the form of the matrix elements in detail in the next section,
 but at the symmetry point they can be shown to have the approximative form
  \begin{equation}
 \begin{split}\label{eq_ElementsFlux}
 M_1^2 &= \sin^2\left(\frac{\varphi_{\rm min}}{2} \right),\\
 M_2^2 &= e^{-2\omega' \varphi_{\rm min}^2},
 \end{split}
 \end{equation}
 where $\varphi_{\rm min}$ is the solution to the transcended equation
 \begin{equation}
  E_L\varphi_{\rm min}-E_J\sin\varphi_{\rm min}=0\, ,
 \end{equation}
  and
  \begin{equation}
   \omega'=\sqrt{\frac{E_L - E_J \cos(\varphi_{\rm min})}{2E_C}} \, .
  \end{equation}
 The decay matrix element is the largest of the discussed qubits.
 Important is also that this qubit has relatively low tunneling resistance $R_{\rm T}$.
 Depending on the interfence effect it can be rather sensitive to quasiparticle tunneling.

 The qubits have various interference angles, as defined in Eq.~(\ref{eq_AngleDefinition}).
 For the qubits that are similar to a harmonic oscillator this angle is given by the position of the local minimum.
 The potentials of the transmon and fluxonium qubits are symmetric around the phase $\varphi_0=0$,
 whereas the potential of the phase qubit is also practically symmetric, for certain $\varphi_0 \neq 0$.
 In the dc-flux qubit (with $\varphi_{\rm ext}=\pi$) the interference angle is given by $\varphi_0=\pi$,
 and the situation is the same for the stronlgy anharmonic phase qubit.
 This provides protection against qubit decay due to non-equilibrium quasiparticles, as then
 the interference between two electron tunneling directions is destructive. However, this protection is only
 partial because for typical qubit splittings $\delta E\approx\Delta/10$ one has $\epsilon_{I}\approx 0.9$ ($<1$).

 The common decay rate (\ref{eq_MainResult}) shows that the two quasiparticle sources produce a similar type of decay rate.
 It is now easy to estimate which source should be dominant in typical experimental conditions.
 A crucial quantity is the dimensionless parameter $\xi_i/\Delta$, defined in Eq.~(\ref{eq_MainDensity}).
The division by the energy gap also accounts for changes in Eq.~(\ref{eq_MainResult}) due to change in the tunneling resistance,
 if the same qubit (with same $E_{\rm J}$)
 is build from, for example, Niobium instead of Aluminum,
 as one has $R_{\rm T}\propto\Delta$.
 For an Aluminum superconductor the parameter is usually measured to have
 a value in the range $\xi_I/\Delta\sim 10^{-8}-10^{-5}$ (non-equilibrium quasiparticles) or  $\xi_{II}/\Delta\sim 10^{-15}-10^{-9}$ (sub-gap density of states).
 This means that for Aluminum non-equilibrium quasiparticles have usually an impact that is several orders of magnitudes larger than
 that of sup-gap states.
 On the other hand, for Niobium using the typically observed Dynes parameter~\cite{Toppari,Pashkin} we obtain values $\xi_{II}/\Delta\sim 10^{-6}-10^{-5}$,
 corresponding to the upper bound values for non-equilibrium quasiparticles in Aluminum.

 In table \ref{tab_comparing_decay_times} we compare decay times for various qubits calculated by
 assuming the lowest reported value for the density of non-equilibrium quasiparticles in Aluminum, given in Ref.~\onlinecite{Pekola}.
 For single-junction qubits (transmon and phase)  we find decay times of the order of ms, which is
 two to three orders of magnitude higher than the best experimental observations.
 For the same qubits but build from Niobium the decay rate, Eq.~(\ref{eq_MainResult}), simplifies to
 \begin{equation}
  \Gamma_1\approx\frac{1}{C R_{\rm T}(\Delta/\Gamma_{\rm D})^2},
 \end{equation}
 being consistent with the result of the classical limit.
 In this case we find decay times of the order of $10$~$\mu$s.

 The single-junction qubit that is the most sensitive to both quasiparticle sources is the dc-flux qubit, for which we obtain
 decay times of order 100~$\mu$s (Al) and 100~$n$s (Nb) (not listed in table \ref{tab_comparing_decay_times}). The value for Aluminum includes the protection
 due to destructive interference which is approximately a factor of $10$.
 All together, in these conditions the Aluminum based single-junction qubits are protected against
 non-equilibrium quasiparticles, most of them due to small $E_{\rm C}/E_{\rm J}$-ratio.


\begin{figure}[t]
 \begin{center}
 \includegraphics[width = 8 cm]{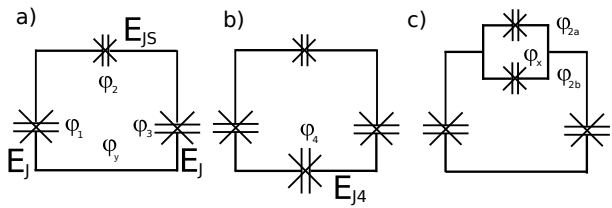}
  \caption{   (a) The original persistent-current flux qubit
              as proposed in Ref.~\onlinecite{VonDelftPersitentFluxQubit}. The p-flux qubit has three junctions, two large junctions
              with Josephson energy $E_{\rm J}$  and phase differences $\varphi_1$ and $\varphi_3$. The smaller junctions has the Josephson energy $E_{\rm JS}$
              and the phase difference $\varphi_2$. Through the loop a flux $\varphi_y$ is applied.
             (b) For technical reasons many flux qubits actually have four junctions. The fourth junction has the Josephson energy
                  $E_{{\rm J}4}$ and the phase difference $\varphi_4$.
                (c) To allow for more tunability often the smaller junction
                    is replace by a SQUID loop consisting of two junctions with phase differences
                    $\varphi_{2a}$, and $\varphi_{2b}$ }\label{fig_The_Flux_qubit}
 \end{center}
\end{figure}


\section{Multi-junction flux qubits}\label{sec_Multi_Junction_Flux_Qubit}

 In this section we generalize our analysis to multi-junction qubits. We consider in detail
 the most common of them, the flux qubit, which comes in many different shapes and forms.
 The original proposal, Fig.~\ref{fig_The_Flux_qubit}(a),  is a 
 qubit with three junctions~\cite{VonDelftPersitentFluxQubit}, but 
 in general it is today used in many variations
 often with several extra junctions. For technical reasons the flux 
 qubits have always been build with a fourth junction, Fig.~\ref{fig_The_Flux_qubit}(b),
 which in most modern flux qubits is of the same size as the two large junctions. 
 In another version of the flux qubit, the small junction
 has been replaced with a SQUID, meaning by a loop with two junctions, Fig.~\ref{fig_The_Flux_qubit}(c).  
 This allows for more tunability. 

 We will start this section with a discussion of the generalization of the decoherence rates to many junctions. 
 Then we will continue with calculating the rates for
 the p-flux qubit with 
 three junctions and then later
 discuss the modification by the fourth junction and the tunable third junction. We end the section with
 the capacitively shunted flux qubit, which by its properties is shown to be similar to an anharmonic oscillator.
  
\subsection{Generalization of the decoherence rates}
 As before we define our total Hamiltonian of the qubit and the quasiparticle environment as 
 \begin{equation}
  H=H_{\rm qubit}+H_{\rm qp}+H_{\rm T},
 \end{equation}
 where $H_{\rm qubit}$ is the Hamiltonian of the flux qubit, coupled via the tunneling Hamiltonian $H_{\rm T}$
 to a quasiparticle environment $H_{\rm qp}$.
 With a total of $n$ junctions we have $H_{\rm T}=\bar H_{\rm T}+\sum_{j=1}^nE_{{\rm J}j}\cos\varphi_j$, where
 the bare tunnel Hamiltonian is
\begin{eqnarray}\label{Ham4}
\bar H_{\rm T}=\sum_{j=1}^{n}t_j\sum_{{\bf kl}\alpha}(c_{j{\bf k}\alpha}^{\dagger}c_{j{\bf l}\alpha}\hat T_j+{\rm h.c.}).\label{eq_QuasiparticleTunneling2}
\end{eqnarray}
Here $c_{j{\bf k}}^{(\dagger)}$ and $c_{j{\bf l}}^{(\dagger)}$ are electron annihilation (creation) operator of the states $\bf k$
and $\bf l$ on the two sides of the junction $j$ with Josephson coupling $E_{{\rm J}j}$, and $\hat T_j$ 
is the corresponding qubit-space charge-transfer operator, given in different cases below. The quasiparticle environments in each of the leads
are assumed to be identical and described by  relations (\ref{eq_GreenDefined}-\ref{broadening2}).

In the leading order the decay rates can be written into a similar form as for the single-junction case, Eq.~(\ref{eq_Gamma12_defined}),
and in the low-energy approximation (section \ref{sec_LowEnergyApproximation}) one gets
\begin{equation}
\begin{split}\label{eq_MainResult3}
\Gamma_{1}^{\rm total}&=\sum_{j=1}^n\Gamma_{1j},\\
\Gamma_{1j}&=\frac{2\xi_iM_{1j}^2}{R_{{\rm T}j}e^2}\left(1+\epsilon_i\cos\varphi_{0j}\right),\\ 
M_{1j}^2&=\left\vert \langle \uparrow |\hat T_j| \downarrow \rangle \right\vert^2.
\end{split}
\end{equation}
Here $R_{{\rm T}j}$ and $\varphi_{0j}$ stand for the tunneling resistance and the interference angle of the junction $j$.
Similar generalization applies also for the pure dephasing rate, Eq.~(\ref{eq_MainResult2}). In the following we discuss the
form of the matrix elements $M_{ij}^2$ and the phases $\varphi_{0j}$  in the considered flux qubit realizations.

 \subsection{The persistent-current flux qubit}

 The first persistent-current flux qubit we consider has three Josephson junctions, see Fig.~\ref{fig_The_Flux_qubit}(a).
 Its Hamiltonian can be presented in the form
 \begin{equation}\label{eq_Full_version_of_persisten_Flux_Qubit_Hamiltonian}
 \begin{split}
  H_{\rm qubit}& =E_C\left(\frac{1}{1+2\alpha_s}N_+^2 + N_-^2\right)\\
    &-E_{\rm J}\left[2\cos(\varphi_+)\cos(\varphi_-)+\alpha\cos(2\varphi_+-\varphi_{\rm ext})\right] .
 \end{split}
 \end{equation}
  Here $E_C$ is the total charging energy of the system, $E_{\rm J}$ is the Josephson energy of the larger junctions and $\alpha=E_{\rm JS}/E_{\rm J}$ is
  a parameter that determines the energy splitting of the qubit. In the following it is assumed that for the corresponding
  capacitive term it holds $\alpha_s=\alpha$.
 The ratio of Josephson
 and charging energy for the p-flux qubit is generally 
 given by $E_J/E_C\approx 50$.
  We have three junctions with phase differences $\varphi_1$, $\varphi_2$ and $\varphi_3$, as illustrated in Fig.~\ref{fig_The_Flux_qubit}(a).
 The phases in the Hamiltonian are then defined by
 \begin{equation}
  \varphi_{\pm}=\frac{1}{2}\left(\varphi_1 \pm \varphi_3 \right)\,\, ,\,\, \varphi_{\rm ext}=\varphi_y\, .
 \end{equation}
 For each dynamical phase we have a conjugate charge variable, $[\varphi_{\pm},N_{\pm}/2]=i$, and
  for each junction we have a tunneling operator given by
 \begin{equation}\label{Tunneling_operators_for_persitent_flux_qubit}
 \begin{split}
  \hat T_{1}&=e^{i(\varphi_+ + \varphi_-)/2}\,\, ,\,\, 
  \hat T_{2}= e^{i(\varphi_{\rm ext}-2\varphi_+)/2} ,\\
  \hat T_{3}&=e^{i(\varphi_+ - \varphi_-)/2} \, .
 \end{split}
 \end{equation}

\subsubsection{Eigenstates}

 The potential of the p-flux qubit is given by
 \begin{equation}
 \begin{split}
  U(\varphi_-,\varphi_+)/E_J &= -2\cos(\varphi_+)\cos(\varphi_-)\\
                             &  -\alpha\cos(2\varphi_+-\varphi_{\rm ext}).
 \end{split}
 \end{equation}
  In the direction of $\varphi_-$ the system behaves similar to a harmonic oscillator that remains in the ground state.
 Therefore we will focus on the dynamics in the $\varphi_+$ direction.
  Choosing $\varphi_{\rm ext}=\pi$ (the symmetry point) and $\varphi_-=0$  we expand $U$ for small $\varphi_+$,
 \begin{equation}\label{eq_U_flux_qubit_approximated_in_lowest_order}
   U/E_J\approx(1 - 2 \alpha) \varphi_+^2 +\left[ \frac{2\alpha}{3} -\frac{1}{12} \right] \varphi_+^4.
 \end{equation}
 We have $\alpha\gtrsim 1/2$.
 This means that the eigenstates of the qubit are formed by symmetric and antisymmetric superpositions
 of the ground state of two wells. These ground states
 are centered around the minima $\pm\varphi_{+,{\rm min}}$,
 \begin{equation}
  \varphi_{+,{\rm min}}=\frac{\sqrt{6}\sqrt{2\alpha-1}}{\sqrt{8\alpha-1}}.
 \end{equation}
 In the left well the eigenstate is of the form
 \begin{equation}
  \langle \varphi_+ |L\rangle = \left(\frac{\omega'}{\pi}\right)^{1/4} \exp\left[-\omega' (\varphi_+ + \varphi_{+,{\rm min}})^2/2\right],
 \end{equation}
 where we defined
 \begin{equation}
  \omega' =\sqrt{(4\alpha^2+1)\frac{E_J}{2E_C}}.
 \end{equation}
 The state in the right well, $|R\rangle$,
 has the same form, just centered at $\varphi_{+,{\rm min}}$. Under the same approximation we used to
  derive the potential given by Eq.~(\ref{eq_U_flux_qubit_approximated_in_lowest_order}) we can now find the coupling strength $t$
  between the two states, which is given by
 \begin{eqnarray}
    \frac{t}{2}&=&\langle R| H_{\rm qubit}|L\rangle\\
         &\approx&\frac{[-2 - 3 \alpha + 2 \alpha^2 - 2\varphi_{+,{\rm min}}^2 (4\alpha^2+1)] E_J}{ 1 + 2 \alpha}\nonumber\\
         & & \times \exp(-\omega'\varphi_{+,{\rm min}}^2)\nonumber.
 \end{eqnarray}
 Here we assumed $\varphi_{\rm ext}=\pi$ and
 consider the contribution of the order of $E_{\rm J}$. There is an additional contribution of the order of $\sqrt{E_{\rm J} E_C}$
  and $E_C$, but they remain small.
  Often the coupling strength has been calculated using the WKB approximation. Our calculations underestimates the coupling strength
  but still gives a good order of magnitude approximation. 
  
  We move away from symmetry like $\varphi_{\rm ext}=\pi-\delta\varphi_{\rm ext}$ and
  in the lowest order of $\delta\varphi_{\rm ext}$ we get
  \begin{equation}
   H_{\rm qubit}\approx H_{\rm qubit}(\varphi_{\rm ext}=\pi)+E_J\left(-2\alpha\varphi_+ +\frac{4\alpha \varphi_+^3}{3} \right)\delta\varphi_{\rm ext}.
  \end{equation}
   If we now compare the energy of the minima with the energy of the minima for $\delta\varphi_{\rm ext}=0$ we find the 
   energy difference
   \begin{equation}
    \delta \epsilon=  E_{\rm J}\frac{12\alpha\sqrt{6}\sqrt{2\alpha -1}}{(8\alpha-1)^{3/2}}\delta\varphi_{\rm ext}.
   \end{equation}
   Now we can write the Hamiltonian in the standard way of a two-state approximation for the states $|L\rangle$ and $|R\rangle$,  
  \begin{equation}
   H_{\rm qubit}\approx
   \frac{1}{2}\left(\begin{array}{cc}
               \delta \epsilon &          t                \\
                       t       & -\delta \epsilon
              \end{array}\right).
  \end{equation}
  The basic form of the states is therefore given by 
  \begin{equation}\label{eq_eigenstates_of_p_flux_qubit}
  \begin{split}
   |\uparrow\rangle &= \cos\left( \frac{\theta}{2} \right)|L\rangle +\sin\left( \frac{\theta}{2} \right) |R\rangle, \\
   |\downarrow\rangle &=  - \sin \left( \frac{\theta}{2} \right) |L\rangle +\cos\left( \frac{\theta}{2} \right)|R\rangle,
  \end{split}
  \end{equation}
  with $\tan \theta=t/\delta \epsilon$.

  Let us now consider an operator of the form
  \begin{equation}
   \hat T=e^{i\varphi_+/2}.
  \end{equation}
   We get in a good approximation
  \begin{equation}
   \langle L | \hat T|L\rangle =e^{-i\varphi_{+,{\rm min}}/2}\,\,
   , \,\, \langle R | \hat T| R\rangle = e^{i\varphi_{+,{\rm min}}/2},
  \end{equation}
   and
   \begin{equation}
     \langle L |\hat T|R\rangle =  \langle R |\hat T|L\rangle =e^{-\omega' \varphi_{+,{\rm min}}^2}.
    \end{equation}
    Using these equations we will be able to calculate the effect of quasiparticle tunneling on the flux qubits.

\subsubsection{Qubit decay}
  To estimate the qubit decay rate we start from the transition element of the operator  
 $\hat T=e^{i\varphi_+/2}$. This can be expressed in the form
  \begin{eqnarray}
   \langle \downarrow| \hat T | \uparrow \rangle = i\sin\theta \sin \left( \frac{ \varphi_{+,{\rm min}} }{2} \right) +\cos\theta e^{-\omega' \varphi_{+,{\rm min}}^2} .
  \end{eqnarray}
  From this general form we can deduce 
  the results for the tunneling operators $\hat T_1$ and $\hat T_3$,
  \begin{equation}\label{eq_PFLuxElement1}
  \begin{split}
  \langle \downarrow| \hat T_1 | \uparrow \rangle &=  \langle \downarrow| \hat T_3 | \uparrow \rangle= i\sin\theta \sin\left( \frac{ \varphi_{+,{\rm min}} }{2} \right) \\
  &+\cos\theta e^{-\omega' \varphi_{+,{\rm min}}^2},
  \end{split}
  \end{equation}
   where we have used the fact that the system stays in the ground state in the direction $\varphi_-$.
   Important here is that for the symmetry point one has $\theta=\pi/2$ and the matrix 
   elements (\ref{eq_PFLuxElement1}) are purely imaginary.
   This is in contrast to the (single-junction)  dc-flux qubit where the element is real. It follows that
   the interference angles in Eq.~(\ref{eq_MainResult3}) are $\varphi_{01}=\varphi_{03}=0$, i.e., one has
   constructive interference instead of destructive.

   The junction that is somewhat different is the second junction, for which
   the matrix element is given by
   \begin{equation}\label{eq_PFLuxElement2}
   \langle \downarrow| \hat T_2 | \uparrow \rangle 
   = \left[-i\sin\theta \sin\varphi_{+, {\rm min} }+\cos\theta e^{-\omega' \varphi_{+, {\rm min} }^2}
       \right] e^{i\varphi_{\rm ext}/2}.
   \end{equation}
   The element (\ref{eq_PFLuxElement2}) has an additional phase factor, $e^{i\varphi_{\rm ext}/2}$,
   and  becomes real at the symmetry point $\varphi_{\rm ext}=\pi$. This corresponds to the case of 
   the dc-flux qubit and destructive interference.

 The interference effect has an important role in determining
 the sensibility to non-equilibrium quasiparticle tunneling.
 As discussed in section \ref{DiscussionForSingleQubits},
 the decay matrix element of this qubit type is the largest,
 but the single junction dc-flux qubit is protected by destructive interference.
 As a result of constructive interference for two junctions of the p-flux qubit, this protection is lost. 
 It follows that even with the smallest observed non-equilibrium quasiparticle density~\cite{Pekola}, corresponding to
 $\xi_I/\Delta\sim 10^{-8}$,
 we obtain a decay time of the order $10$~$\mu$s. This result coincides with recent experiments \cite{NakamuraSuperFux}.

 Results for Niobium p-flux qubits is similar to dc-flux qubits:
 using typical subgap densities\cite{Toppari}
 one obtains decay times of the order $100$~$n$s, being also similar to recent experiments 
 \cite{MDOliver}. These results are compared with
 the ones of other qubit types in table~\ref{tab_comparing_decay_times}.

  \subsubsection{Dephasing}
  
  The dephasing is characterized by the diagonal matrix elements of $\hat T_j$, which can be shown to have the form
   \begin{eqnarray}
   \begin{split}
    \langle \uparrow|T_1|\uparrow\rangle & =  \langle \uparrow|T_3|\uparrow\rangle= -i\cos\theta\sin\left( \frac{\varphi_{+,{\rm min}}}{2}\right) \\
    & +\sin\theta e^{-\omega'\varphi_{+,{\rm min}}^2}+\cos\left(\frac{\varphi_{+,{\rm min}}}{2}\right).
   \end{split}
\\
   \begin{split}
    \langle \downarrow|T_1|\downarrow\rangle & =  \langle \downarrow|T_3|\downarrow\rangle= +i\cos\theta\sin\left( \frac{\varphi_{+,{\rm min}}}{2}\right) \\
    & -\sin\theta e^{-\omega'\varphi_{+,{\rm min}}^2}+\cos\left(\frac{\varphi_{+,{\rm min}}}{2}\right).
   \end{split}
   \end{eqnarray}
    The second junction has again an extra phase dependence,
   \begin{equation}\label{eq_PFluxDephasing2}
   \begin{split}
   & \langle \updownarrow|T_2|\updownarrow\rangle = e^{i\varphi_{\rm ext}/2}\times \\ 
   &\left[\pm i\cos\theta\sin\left( \varphi_{+,{\rm min}}\right)\pm \sin\theta e^{-\omega'\varphi_{+,{\rm min}}^2} +\cos\left( \varphi_{+,{\rm min}}\right) \right].
   \end{split}
   \end{equation}
   The presence of a term that does not change sign when changing the state leads effectively to a different angle $\varphi_{0^*}$
   for the two states $\vert\uparrow\rangle$ and $\vert\downarrow\rangle$, and deviations from result (19). However, near the symmetry point
   such corrections stay small and one can neglect this effect.
   In this region the main contribution to pure dephasing comes from the junction 2.
   Using Eq.~(\ref{eq_PFluxDephasing2}) one obtains then for the relevant dephasing matrix element
    \begin{equation}
   \left\vert \langle \uparrow|T_2|\uparrow\rangle - \langle \downarrow|T_2|\downarrow\rangle \right\vert^2 = e^{-2\omega'\varphi_{+,{\rm min}}^2}.
   \end{equation}  
   This is usually much smaller than the decay element. Combined with the discussion of section \ref{DiscussionForSingleQubits}, this  means that at the symmetry point
   the dephasing is limited only by the qubit decay.

\subsection{The four-junction flux Qubit}
 
  For technical reasons the p-flux qubit is in fact always build with four junctions
  instead of the necessary three. The fourth junction has a charging energy $E_{C4}$ and a Josephson
  energy $E_{{\rm J}4}$. We can assume $E_{{\rm J}4}\gg E_{C4}$. The additional junction adds another dimension to the problem, and
  we can approximate states along this dimension as states of a harmonic oscillator. Within this approximation
  the total Hamiltonian can be written as 
  \begin{equation}\label{eq_Hamiltonian_for_four_junction_flux_qubit_in_JC_form}
  \begin{split}
   H_{\rm qubit} &= \frac{1}{2}\delta E\tau_z+g\left(m_1\tau_x+m_2\tau_z\right)(a^{\dag}+a) \\
                 & +\omega a^{\dag}a,
  \end{split}
  \end{equation}
  where $\tau_i$ are the Pauli matrices acting on the eigenstates of the
  p-flux qubit (\ref{eq_eigenstates_of_p_flux_qubit}), with the energy splitting $\delta E=\sqrt{\delta \epsilon^2+t^2}$.
  These states are coupled with the coupling strength $g=E_J\alpha(2E_{C4}/E_{J4})^{1/4}$ to the fourth junction, modeled
   as a harmonic oscillator with frequency $\omega=\sqrt{8 E_{C4} E_{J4}}$. The coupling is determined by the following matrix elements,
  \begin{equation}
  \begin{split}
    m_1 &= \langle \uparrow |\sin\left( 2\varphi_+ -\varphi_{\rm ext} \right)|\downarrow\rangle, \\
   2m_2 &= \langle \uparrow | \sin\left( 2\varphi_+ -\varphi_{\rm ext}  \right)|\uparrow \rangle 
             -\langle \downarrow | \sin\left( 2\varphi_+ -\varphi_{\rm ext}  \right)|\downarrow \rangle\, . 
  \end{split}
  \end{equation}
 If the  fourth junction is large, 
 $E_{{\rm J}4}\rightarrow \infty$, the coupling goes to zero. In this case the fourth junction is decoupled and always
 stays in the ground state.  However, in many flux qubits the fourth junction has the
 Josephson energy $E_{{\rm J}4}\approx E_{\rm J}$. In this case the coupling is strong,
 $g \approx \omega $.  

 The tunneling operator for the fourth junction is simply given by $T_4=e^{i\varphi_4/2}$.
 Close to the symmetry point we have $m_2\approx 0$ and additionally we assume $\omega \gg \delta E$.
 In this case we can approximate the matrix element in the lowest order of $\left(2E_{C4}/E_{{\rm J}4}\right)^{1/4}$ by
 \begin{equation}\label{eq_T4_in_maximum_approximation}
     \langle g| T_4| e\rangle=-i\left(\frac{ 2E_{C4}}{E_{{\rm J}4}}\right)^{1/4} \frac{g m_1}{\sqrt{\omega^2+4 g^2 m_1^2}},
 \end{equation}
 where we have labeled the two lowest eigenstates of the four-junction flux qubit as $\vert g\rangle$ and $\vert e\rangle$.  
 In the limit of weak coupling $E_{{\rm J}4}\gg E_{{\rm J}}$, we find $\omega\gg g m_1$. Then the matrix element
 simplifies to $\langle g| T_4| e\rangle=-i\alpha E_{\rm J} m_1/8 E_{{\rm J}4}$.
 In general we have $E_{C4}/E_{{\rm J}4}\ll 1$ and therefore transitions due to tunneling across the fourth junction
 are well suppressed compared to tunneling across the other junctions.
 Similarly we can calculate the tunneling through the third junction with the tunneling operator $T_{3}=e^{i(\phi_+ - \phi_-)/2}$
 and, using the same approximation as for Eq.~(\ref{eq_T4_in_maximum_approximation}), we get
 \begin{equation}
   \langle g| T_3| e\rangle\approx \langle \downarrow | T_3 | \uparrow\rangle\, .
 \end{equation}
 In conclusion
 we can say that the fourth junction should have little effect on
 the overall decoherence rate of the p-flux qubit.

   \subsection{Tunable gap flux qubit}

  In the tunable gap flux qubit the small junction is replaced by a SQUID, see Fig.~\ref{fig_The_Flux_qubit}(c).
  In this case the parameter $\alpha$ can be tuned by changing the field $\varphi_x$. In
  Hamiltonian~(\ref{eq_Full_version_of_persisten_Flux_Qubit_Hamiltonian})
  this changes 
  \begin{equation}
  \begin{split}
  \alpha &=\alpha_0\cos\left(\frac{\varphi_x}{2}\right),\\
  \varphi_{\rm ext} &=\varphi_y+\frac{\varphi_x}{2}.
  \end{split}
  \end{equation}
   The four tunneling operators  are given by
  \begin{equation}
  \begin{split}
  T_{1}&=e^{i(\varphi_+ + \varphi_-)/2}\,\, ,\,\,\,\,\,\,\,\,\,\,\,\,\,\,\,\,\,\,\,  T_{2a}=e^{i(\varphi_{\rm ext}+\varphi_x/2-2\varphi_+ )/2},  \\
  T_{2b}&= e^{i(\varphi_{\rm ext}-\varphi_x/2-2\varphi_+ )/2}  \,\, ,\,\,
  T_{3}=e^{i(\varphi_+ - \varphi_-)/2},
  \end{split}
  \end{equation}
  where $T_{i}$ corresponds to the junction with phase $\varphi_i$.  
  Since the eigenstates are the same as for the p-flux qubit,
  it is straightforward to calculate the tunneling matrix elements. The decay rates for quasiparticle tunneling through the large junctions
  are the same as for the standard persistent-current flux qubit and we therefore get the similar overall rates. 
  The only difference is that the tunneling across the small junctions, that form the SQUID, have now different type of
  interference effect if $\varphi_x\neq 0$.

\subsection{Capacitively shunted Flux qubit}

 Another version of the same
 qubit, the capacitively-shunted flux qubit \cite{Capacativelyshuntedfluxqubti} 
 has the same Hamiltonian as the persistent-current flux qubit (\ref{eq_Full_version_of_persisten_Flux_Qubit_Hamiltonian}).
 The major difference is that it is operated in the regime $\alpha \leq 1/2$ and
 the charging energy along the dimension of the double well potential is significantly reduced by introducing a shunt capacitance,
 $\alpha_s\gg \alpha$. This qubit has excellent coherence times and still preserves relatively large anharmonicity.
  We will discuss the system at the symmetry point
 $\varphi_{\rm ext}=\pi$ and for the case where it deviates most strongly from a harmonic oscillator, $\alpha=1/2$.
 After this we are able to generalize this result to other parameter regimes.
 
 Under the conditions we specified above we can write the Hamiltonian as 
 \begin{equation}
   H_{\rm qubit}({\varphi_{\rm ext}=\pi,\alpha=1/2})\approx E_C' N_{+}^2 +\frac{1}{3}E_{\rm J} \varphi_+^4,
 \end{equation}
 with $E_C'=E_C/(1+2\alpha+C_s/C_J)$.
 We note that the Hamiltonian is similar to the one of the strongly anharmonic phase qubit \cite{Zorin}.   
 We find the eigenstates to be
 \begin{eqnarray}
     \langle\varphi_+|\downarrow\rangle  &=& \left(\frac{\omega'}{\pi}\right)^{1/4} e^{-\omega' \varphi_+^2}, \\
     \langle \varphi_+ |\uparrow\rangle    &=& \left(\frac{\omega'}{\pi}\right)^{1/4} \sqrt{2\omega'}\,\varphi_+\, e^{-\omega'\varphi_+^2/2},
 \end{eqnarray}
 with $\omega'=\left(3 E_{\rm J}/16 E_C'\right)^{1/4}$.
 It is now simple to calculate the matrix elements for the tunneling operators (\ref{Tunneling_operators_for_persitent_flux_qubit}),
 \begin{equation}
      \langle \downarrow | \hat T_1 |\uparrow \rangle= \langle \downarrow |\hat T_3 |\uparrow \rangle \approx \frac{i}{2}\left(\frac{2E_C'}{3E_J}\right)^{1/6},
 \end{equation}
 and for $\hat T_2$ one obtains again the additional phase factor $e^{i\varphi_{\rm ext}/2}=i$.
 The energy splitting between the eigenstates is given in a good approximation by $ \delta E=2(12 {E_C'}^{2} E_J)^{1/3}$.
 Using this result we can rewrite the matrix element as $\langle \downarrow |\hat T_1 |\uparrow\rangle= \sqrt{E_C'/\delta E} $.
 This is a well known result for a harmonic oscillator and is therefore also valid for the complete
 relevant parameter regime of the capacitively shunted flux qubit.
 The interference angles $\varphi_{0j}$ of the three junctions are the same as for the p-flux qubit at the symmetry point:
 In the case of non-equilibrium quasiparticles and qubit decay the junctions 1 and 3 have
 constructive ($\varphi_{01}=\varphi_{03}=0$) and the junction 2 has destructive interference ($\varphi_{02}=\pi$). Vice versa for the dephasing.
      
 As mentioned, the c-flux qubit is very similar to the strongly anharmonic phase qubit,
 with the difference that the latter qubit has only a single junction, which corresponds to the second junction 
 of the c-flux qubit. As this junction shows destructive interference against qubit decay due to non-equilibrium quasiparticles,
 the strongly anharmonic phase qubit is better protected against this decoherence source.
 However, the matrix elements of the qubits are generally similar to that of harmonic oscillators, and this means that the c-flux qubit
 is relatively well protected against quasiparticle tunneling, too.
 In table \ref{tab_comparing_decay_times} we compare the qubit 
 decay times between the capacitively shunted flux and the other discussed qubits.

\section{Conclusion}\label{Sec_Conclusion}

In this work we analyzed decoherence in  superconducting qubits due to
quasiparticle tunneling. We considered
two types of sources of quasiparticle tunneling relevant for low temperatures and low qubit energies: non-equilibrium quasiparticles and sub-gap density of states.
Using typically observed values for their densities
we estimated the resulting qubit decay times and compared them with the experimentally measured ones.
In the best case scenario, i.e., with the lowest reported quasiparticle densities,
we showed that only the decay times of persistent-current flux qubits were similar to recent experiments.
The multi-junction flux qubits have achieved recently excellent coherence times and have 
 the remarkable feature of extremely large anharmonicity
 which makes it the best approximation to a real two level 
 system of all superconducting qubits.
 However, the presented analysis shows that such qubits are also very fragile against quasiparticle tunneling induced decoherence.
To protect the qubits against creation of non-equilibrium quasiparticles, for example,
 a careful isolation from the nearby environments should be realized~\cite{Pekola,MartinisStrayLight}.

\section*{Acknowledgments}

We thank A.~Heimes, W.~D.~Oliver, P.~Kotetes, G.~Sch\"on, J. Liesenfeld and G.~Johansson for useful discussions.
This work was supported by the CFN of DFG
and the U.S. ARO under Contract No. W911NF-09-1-0336.

\section*{Appendix}

\subsection{Derivation of the decoherence rates}

Here we formulate the qubit decay rate as a function of
electron correlation (Green's) functions of the leads.
Our aim is to consider the response of the superconductor as a sum over all
energy states, rather than consider the response of single states.
By this we can model situations where
the superconductor quasiparticle density of states at a given energy cannot be mapped back into certain
quasiparticle energy in the normal state, for example, due to energy-level broadening effects.

We consider a single Josephson junction qubit and write the time
evolution of the reduced density matrix, when interacting with
quasiparticle environment, generally as
\begin{equation}
\begin{split}
\dot\rho_{mn}(t)&=i(E_n-E_m)\frac{\rho_{mn}}{\hbar}\\
&+\sum_{ab}\int_{t_0}^tdt'\sigma^{a\rightarrow m}_{b\rightarrow n}(t-t')\rho_{ab}(t').
\end{split}
\end{equation}
Here the (generalized transition rate) tensor $\sigma_{b\rightarrow n}^{a\rightarrow m}$ includes the effect of qubit-quasiparticle interaction.
Its calculation is similar to the case of metallic reservoirs~\cite{SchoellerSchon} with a difference that
when tracing out the environment also two annihilation (creation) operators $c_k^{(\dagger)}$
can also contract to pairs. Their contribution leads to the pair density $p$, whereas diagonal contributions, such as $c_k^{\dagger}c_k$,
lead to the density of states $n$.


In the leading order we obtain for the transition from the state $\vert i\rangle$ to $\vert f\rangle$
\begin{equation}
\begin{split}
&\sigma_{i\rightarrow f}^{i\rightarrow f}=2{\rm Re}\left\{\lim_{s\rightarrow 0} \int_{-\infty}^tdt'e^{-(t-t')s} \bar\sigma(t-t')  \right\},\\
&\bar\sigma(t-t')=\left\langle  \langle i\vert \bar H_{\rm T}(t) \vert f\rangle\right   \langle f\vert \bar H_{\rm T} (t') \vert i\rangle   \rangle
\end{split}
\end{equation}
The trace over the initial distribution of quasiparticles leads to four nonvanishing
contributions
\begin{equation}
\bar\sigma(t)=\frac{t^2}{\hbar^2}e^{i\delta \omega t}(a+b+c+d),
\end{equation}
where $\delta \omega=\delta E/\hbar$ and
\begin{equation}
\begin{split}
a&= \vert\langle f\vert \hat T \vert i\rangle\vert^2  \sum_{kl\alpha} G^{>}_{\alpha\alpha}({\bf k},t-t')G^{<}_{\alpha\alpha}({\bf l},t'-t), \\
b&= \vert\langle f\vert \hat T^{\dagger} \vert i\rangle\vert^2 \sum_{kl\alpha} G^{<}_{\alpha\alpha}({\bf k},t'-t)G^{>}_{\alpha\alpha}({\bf l},t-t'), \\
c&=-\langle i\vert \hat T \vert f\rangle \langle f\vert \hat T \vert i\rangle \sum_{kl\alpha\beta}F^{>}_{\alpha\beta}({\bf k},t-t')F^{<}_{\alpha\beta}({\bf l},t'-t), \\
d&=-\langle i\vert \hat T^{\dagger} \vert f\rangle \langle f\vert \hat T^{\dagger} \vert i\rangle \sum_{kl\alpha\beta}F^{<}_{\alpha\beta}({\bf k},t'-t)F_{\alpha\beta}^{>}({\bf l},t-t') .
\end{split}
\end{equation}
The correlation functions $G$ and $F$ are defined
in Eq.~(\ref{eq_GreenDefined}).
We define their Fourier transforms as
\begin{equation}
G^{\lessgtr}({\bf k},t)=\frac{1}{2\pi}\int_{-\infty}^{\infty}d\omega e^{-i\omega t}G^{\lessgtr}({\bf k},\omega).
\end{equation}
The Fourier tranforms are related to the spectral densities as in Eq.~(\ref{eq_FourierTransforms}).
In our analysis we have omitted the overall phase appearing usually in $F$-functions, as it is treated
in the qubit part. In compared to Refs.~\onlinecite{Barone,ABRelation} our definition of
$F^{>}$ corresponds to $\tilde F^{>}$ (and
$F^{<}$ to $F^{<}$).

Provided by the assumption of constant tunneling amplitude $t$ we
can now write the contributions to the transition rate
coming from the different terms as
\begin{widetext}
\begin{equation}
a\rightarrow\frac{D^2t^2 }{\hbar^2}\vert\langle f\vert \hat T \vert i\rangle\vert ^2 \lim_{s\rightarrow 0}{\rm Re}\left\{e^{i\delta\omega t'} \int_{-\infty}^tdt'e^{-(t-t')s}\int_{-\infty}^{\infty}d\omega\int_{-\infty}^{\infty}d\omega'n(\omega)n(\omega')f^+(\omega)f^-(\omega')e^{i(\omega'-\omega)t'} \right\}.
\end{equation}
We have defined the normalized density of states $n$ as in Eq.~(\ref{eq_Densities}).
Using the relation $\lim_{s\rightarrow 0}\int_{-\infty}^0e^{i\omega t+st}=\pi\delta(\omega)-iP(1/(\omega))$, one obtains
\begin{equation}
a\rightarrow D^2\pi t^2 \vert\langle f\vert \hat T \vert i\rangle\vert ^2 \int_{-\infty}^{\infty}d\omega \int_{-\infty}^{\infty}d\omega' f^+(\omega)f^-(\omega')n(\omega)n(\omega')\delta(\omega-\omega'-\delta\omega).
\end{equation}
\end{widetext}
Note the symmetry of the equation around $\omega=0$: contribution of negative energies ($\omega,\omega'<0$) gives similar contribution as positive energies ($\omega,\omega'>0$).
As all processes described by $a$ correspond to given electron tunneling direction, negative energies correspond to opposite
tunneling direction of the quasiparticle.
The contribution from $b$ follows the same calculation and is similar to $a$, with a difference that the qubit matrix element
is changed to $\vert\langle f\vert \hat T^{\dagger} \vert i\rangle\vert ^2$.
It corresponds to tunneling of an electron to the opposite direction.

The contribution from $c$ and $d$ exists only if a factor
$\langle i\vert \hat T \vert f\rangle\langle f\vert \hat T \vert i\rangle=\langle i\vert \hat T^{\dagger} \vert f\rangle ( \langle f\vert \hat T^{\dagger} \vert i\rangle )^*$
is finite. It means that it exists if two electron tunneling directions can lead to the same final state, in a single quasiparticle tunneling process. Then the processes interfere.  The contribution from $c$ is similar as before, but replaces $n$ functions by $p$ functions,
\begin{widetext}
\begin{equation}
c\rightarrow -\frac{D^2\pi t^2}{\hbar^2} \vert\langle f\vert \hat T \vert i\rangle\vert ^2 {\rm Re} \left\{ e^{i\varphi_0}\int_{-\infty}^{\infty}d\omega \int_{-\infty}^{\infty}d\omega' f^+(\omega)f^-(\omega')p(\omega)p(\omega') \left[  \delta(\omega-\omega'-\omega_{fi}) -i\frac{P}{\omega-\omega'-\omega_{fi}} \right] \right\},
\end{equation}
\end{widetext}
where we have used the fact that for the considered qubits $\vert\langle f\vert \hat T \vert i\rangle\vert=\vert\langle f\vert \hat T^{\dagger}\vert i\rangle\vert$ and then defined
$\langle i\vert \hat T \vert f\rangle\langle f\vert \hat T \vert i\rangle=-e^{i\varphi_0} \vert\langle f\vert \hat T \vert i\rangle\vert ^2$. Noticing, that the contribution from $d$ is equal to contribution from $c$, except with an opposite phase factor $e^{-i\varphi_0}$,
one obtains that the principal value part gives no contribution. Therefore one gets
\begin{widetext}
\begin{equation}
c+d\rightarrow \frac{2D^2\pi t^2}{\hbar^2} \vert\langle f\vert \hat T \vert i\rangle\vert ^2  \cos\varphi_0\int_{-\infty}^{\infty}d\omega \int_{-\infty}^{\infty}d\omega' f^+(\omega)f^-(\omega')p(\omega)p(\omega') \delta(\omega-\omega'-\omega_{fi}).
\end{equation}
We can write now
\begin{equation}
a+b+c+d\rightarrow\frac{2\pi t^2}{\hbar^2} \vert\langle f\vert \hat T \vert i\rangle\vert ^2 \int_{-\infty}^{\infty}d\omega \int_{-\infty}^{\infty}d\omega' f^+(\omega)f^-(\omega') \left[ n(\omega)n(\omega)+\cos\varphi_0p(\omega)p(\omega') \right] \delta(\omega-\omega'-\omega_{fi}),
\end{equation}
\end{widetext}
leading to equation (\ref{eq_Gamma12_defined}).
This expression is similar to equation (9) in Ref.~\onlinecite{ABRelation}, giving also insight to the interpretation of the different terms.
The main term that is missing, when compared to Ref.~\onlinecite{ABRelation}, is the coherent Josephson term ($\sin\varphi$-term),
as it does not contribute in the process considered here. (This term is
treated exactly in the qubit Hamiltonian). Such coherent terms would
contribute through the principal value integration. The dissipative terms are similar. Especially,
the famous $\cos\varphi$-term corresponds to interference between electron and hole-like tunneling in our calculation, described by terms $c$ and $d$.
A new type of measurement of this term with a superconducting charge qubit has been considered in Ref.~\onlinecite{JuhaQPTunnelingPRB}.

\subsection{Coherent terms: Parameter renormalization}
Taking into account the tunneling Hamiltonian $H_{\rm T}$ by perturbation theory includes not only
incoherent processes but also coherent terms leading to renormalization of the qubit parameters.
Their contribution is usually small but can depend, for example, on the non-equilibrium quasiparticle density.
This has been studied in detail in Ref.~\onlinecite{GlazmanDevoretQPTheoryLong}.

The Josephson coupling $E_{\rm J}$ in Eq.~(2) is calculated as the expectation value over the quasiparticle distribution of the operator
\begin{equation}
-\frac{E_{\rm J}}{2}= \left\langle\langle N+2 \vert \bar H_{\rm T}\frac{1}{H_{\rm qp}} \bar H_{\rm T}\vert N\rangle\right\rangle.
\end{equation}
Here the quasiparticle (BCS) Hamiltonian has no dependence on the charge number $N$.
Using fermi distributions for the excitation occupation probabilities one obtains the famous Ambegaokar-Baratoff relation
for $E_{\rm J}$. This corresponds to the usual choice of the Josephson coupling.
In addition, there exists a similar second-order contribution describing reactive behaviour of quasiparticles
(usually referred to as the capacitance renomalization)
\begin{equation}
H_{\rm C'}=\sum_N \vert N\rangle\langle N\vert \bar H_{\rm T}\frac{1}{H_{\rm qp}}\bar H_{\rm T}\vert N\rangle\langle N\vert.
\end{equation}
To calculate this one can introduce a cut-off function $D(E)=1/[1+(E/E_{\rm co})^2]$, where $E_{\rm co}$ is larger than any of the relevant energy scales in the system introduced to avoid divergence of the energy corrections.
At this point this term produces a constant energy shift for all $N$ and therefore does not bring anything new on the system.


Insted of including these terms into the time-dependent problem, it is easier to
estimate their effect by considering energy-level changes using
time-independent perturbation theory.
Using the second-order theory again, but now for the modified tunneling Hamiltonian $H_{\rm T}$,
one obtains a correction to the energy of the state $i$
\begin{equation}\label{correction}
\begin{split}
\delta H_{ii}^{{\rm J}}&=E_{\rm J}\langle i\vert \cos\varphi\vert i\rangle\\
&+ \left\langle\langle i\vert \bar H_{\rm T}\left[\frac{1}{E_i-H_{\rm qp}-H_{\rm qubit}}\right]\bar H_{\rm T}\vert i\rangle\right\rangle,
\end{split}
\end{equation}
where $E_i$ is the energy of the qubit states $\vert i\rangle$.
At this point we see that, for example, the Ambegaokar-Baratoff result is a good approximation as long as the approximative Josephson term $E_{\rm J}\cos\varphi$
and the corresponding term coming from the second term on the right-hand side of Eq.~(\ref{correction}) almost cancel each other.
It occurs if the transition probabilities to virtual
states $\vert v\rangle$ ($v\neq i$) are small, or if the energies of the virtual states stay small $E_v\ll\Delta$.
In the opposite case extra anharmonicity (non-constant energy-level spacing) could emerge due to this correction, as the
effective tunneling coupling can become different for different states.

\subsection{Decoherence due to higher-order effects: Andreev tunneling}
The decoherence processes $I$ and $II$ correspond to leading order (incoherent) tunneling effects in the system.
The most important process in  higher orders is usually the Andreev tunneling. This involves, for example,  tunneling of two electrons from one side,
with leaving two excitation behind, to form a Cooper pair on the other side. Such processes dominate the sub-gap conductance of
normal metal-insulator-superconductor junctions for ideal BCS state on the superconductor side.
However, for the case of perfect SIS junctions no such process exist until the energy $2\Delta$ is somehow provided for creation of two excitations.
The process appears as a step-like behaviour in the current-voltage characteristics of single JJs nearby $eV=\Delta$.
In experiments the height of the step is usually considerably higher than obtained by theory. The reasons for this
are still unclear~\cite{Pinholes}.

For the case of superconducting qubits no energy is available to create two excitations, unless sub-gap density exists at one side of the junction.
Through creation of excitations into subgap region with creating a Cooper-pair on the other side of the junction one obtains a possible contribution
to the decoherence. However, a simple analysis indicates that the rates for such processes are proportional to $(\Gamma/\Delta)^2/N$, where $N\gg 1$ is the effective number of
parallel tunneling channels. It results that contribution from this process should be much smaller than of the process $II$ considered in this
paper.
We conclude that the contribution from higher-order tunneling effects to decoherence in superconducting qubits should stay small.

\end{document}